# From experimentation to engagement: on the paradox of participatory AI and power in contexts of forced displacement and humanitarian crises

Stella Suge

Executive Director, FilmAid Kenya

Sarah W. Spencer

Nyalleng Moorosi

Senior Researcher, The Distributed AI Research Institute (DAIR)

Helen McElhinney

Executive Director, The CDAC Network

Geoff Loane

Chair, The CDAC Network

Sue Black

Professor of Computer Science and Technology Evangelist, Durham University

## Abstract

Across the Global North, calls for participatory artificial intelligence (AI) to improve the responsible, safe, and ethical use of AI have increased, particularly efforts that engage citizens and communities whose well-being and safety may be directly impacted by AI and other algorithmic tools. These initiatives include surveys, community consultations, citizens' councils and assemblies, and co-designing AI models and projects. Far fewer efforts, however, have been made in the Global South, particularly in contexts related to humanitarian crises and forced displacement, where the deployment of AI and algorithmic tools is accelerating. In this paper, we critically examine participatory AI methods and their limitations in these contexts and explore the opinions and perceptions of AI held by displaced and crisis-affected communities. Based on a pilot exercise with communities living in Kakuma Refugee Camp in northwestern Kenya, we find important limitations in some participatory AI approaches which, if used in humanitarian contexts, could increase risks of so-called "participation washing" and algorithmic harm. We argue that these risks are not predominantly driven by varying levels of understanding and awareness of AI but more closely linked to the fundamental power dynamics embedded within the humanitarian sector - between humanitarian aid recipients, service providers, donor governments, and host nations - as well as the power differentials and incentives that exist between AI companies and humanitarian actors. These structural conditions make the case not only for more rigorous participatory methods, but for independent governance architecture capable of holding humanitarian AI to account.

*CCS CONCEPTS • Human-centered computing—Human computer interaction (HCI)—HCI design and evaluation methods—Field studies • Social and professional topics—User characteristics—Cultural characteristics • Applied computing—Computers in other domains*

*¹ The authors of this paper have been listed in reverse alphabetical order to acknowledge the equal value of all our contributions to this work.*

# 1 INTRODUCTION

Across the globe, technology leaders, academics, governments, and multilateral agencies seeking to improve AI governance are exploring ways in which to meaningfully engage the public in the development and/or the use of AI, driven by the belief that inclusivity and transparency are essential for building public trust in AI systems. Public participation in AI development and deployment allows for diverse perspectives to be considered, particularly from those who are directly impacted by AI systems and their outputs. Participatory AI can also help to identify potential risks, biases, and societal impacts that might otherwise be overlooked.

Many argue that soliciting and acting upon the views and recommendations of those directly impacted by AI is particularly important in high-stakes contexts, such as situations where civil or human rights or humanitarian law are under threat and/or where the use of AI directly affects the well-being of individuals and communities. This includes their access to life-protecting humanitarian aid, public and social services as well as the collection and retention of personally identifiable information (PII), such as biometric information.

Nowhere, arguably, could the stakes be higher than in humanitarian crises and contexts,² where communities and individuals have been forced to flee their homes as a result of conflict, disasters, or persecution. Some estimates suggest that over 180 million people across more than 70 countries currently need lifesaving aid and support. Humanitarian agencies are mandated with delivering this support in conflicts, crises, and emergencies. To improve the efficiency, reach, and impact of their operations, many agencies are increasingly designing and piloting AI solutions. These agencies and their donors equally have long-standing commitments to meaningfully engage communities in the design, implementation, monitoring, and evaluation of humanitarian interventions. Yet, it seems only a handful of humanitarian actors are actively seeking the views of and consulting with the communities and individuals who will be directly impacted—negatively and positively—by these humanitarian AI solutions.

Deepening engagement on AI with crisis-affected and/or displaced communities is not only ethical and in-line with humanitarian actors' existing commitments related to community participation, accountability, and shifting power to the Global South, but could contribute to more effective and responsible AI solutions and policy. Building on an earlier proof-of-concept project to consult with both residents of Kakuma town and refugee camps, the CDAC Network and FilmAid Kenya designed and piloted an AI public participation and community consultation exercise for displaced and crisis-affected communities. Through this exercise, we sought to: (1) pilot public participation methods and the extent to which they generated actionable recommendations and opinions about the use of AI in humanitarian contexts, and (2) surface views and opinions from forcibly displaced and crisis-affected communities on AI, including their general perception and awareness of AI, its benefits and risks, and how AI should be governed. This also included identifying measures these communities believed would improve trust and accountability for AI-enabled projects in humanitarian contexts and also which actors were perceived as more trustworthy with regards to AI deployment and governance.

In this paper, we argue that, like other people directly impacted by AI and algorithmic outputs, refugees and crisis-affected communities have a right to participate in the design and deployment of AI-powered solutions, and agencies have an obligation to consult them. Without community engagement, AI innovations risk becoming experimental and extractive. The asymmetries of power between humanitarian aid actors and those they are charged with serving and supporting—remain a critical obstacle to meaningful community engagement and participatory AI. To avoid experimentation and tokenistic engagement, those deploying AI solutions in humanitarian crises must commit to long-term, deep, and purposeful engagement methods that account for these power dynamics and transfer knowledge about AI and agency to crisis-affected communities.

*[2] For the purposes of this paper, we define humanitarian action as interventions designed to save lives, alleviate suffering, and maintain human dignity during and in the aftermath of man-made crises and natural disasters, as well as to prevent and strengthen preparedness for the occurrence of these events. Humanitarian action includes the protection of civilians and those no longer taking part in hostilities, and the provision of food, water and sanitation, shelter, health services and other items of assistance, undertaken for the benefit of affected populations and to facilitate the return to normal life.*

## 2 RELATED WORK

### 2.1 Community engagement and participatory methods in humanitarian interventions

The theoretical foundations and inclusion of community participation in humanitarian response have evolved significantly since their origins in the 1970s. Early participatory approaches in humanitarian action drew on conceptual frameworks conceived within international development theory, including Paulo Freire's seminal work on critical pedagogy and consciousness-raising, which emphasized agency among marginalized communities. Though developed outside the development sector, Arnstein's "ladder of participation" also provided an influential framework for understanding graduated levels of community engagement and has, in recent years, influenced humanitarian action. Chambers' pioneering work in the 1980s marked a crucial shift in development thinking by challenging prevailing top-down conventions and advocating for local knowledge and community-led analysis, critically examined by Barbara Harrel-Bond in refugee settings. Early efforts to integrate participatory approaches into disaster response included the Capacities and Vulnerabilities Analysis framework. This theoretical foundation significantly influenced how humanitarian organizations conceptualized community engagement during crises, though the institutionalization of humanitarian aid in the 1990s often failed to effectively integrate local knowledge systems. Chambers critically examined power relations in development practice later with others specifically demonstrating the risk of participatory methods reinforcing power structures and becoming merely tokenistic.

The humanitarian sector's formal adoption of participatory approaches is evidenced through various institutional commitments, from the Code of Conduct for the International Red Cross and Red Crescent Movement and NGOs in Disaster Relief to the Core Humanitarian Standard (CHS) and the "Participation Revolution" launched at the 2016 World Humanitarian Summit. At the community level, humanitarian actors have established feedback mechanisms and accountability measures that seek to translate these commitments—including the nine commitments contained within the CHS—into concrete action.

While important, these efforts expose fundamental tensions and a critical dichotomy between "purist" and "instrumentalist" approaches to participation. Purist perspectives emphasize communities' fundamental right to agency and view participation as inherently valuable. By contrast, instrumentalist approaches frame community involvement primarily as a means to improving program effectiveness and efficiency. More recent scholarship has highlighted additional barriers to meaningful participation. Feminist scholars provided crucial critiques of participatory approaches, highlighting the failure to properly examine gender power dynamics. Modes of measurement and assessment have also been found to privilege international actors' perspectives over local voices. For example, asking people whether they wish to opt out of a system critical to their survival reveals a fallacy of consent in humanitarian settings.

On top of these important critiques, a significant gap persists between rhetorical and normative commitments to participation and the operational realities of humanitarian action. A recent initiative by the UN's most senior humanitarian representative acknowledged systematic failures in community engagement: "There is a widespread impression amongst humanitarians that the system is constantly listening to affected populations, but there is little evidence that this is in fact the case." This implementation gap reflects deeper structural challenges linked to the political economy of the humanitarian system that persist, where the incentives of humanitarian agencies are often aligned with donor priorities rather than community-centered approaches, despite innovative advances in participatory methods.

## 2.2 Public participation in AI as a form of governance and accountability

In the last several years, calls for civic engagement and public participation in both the design of AI models and their use have increased. As with participatory humanitarian action, public participation efforts linked to AI development and deployment is typically framed as both a key component of AI governance and accountability—a means for ensuring the safe, responsible, and ethical design and use of AI—as well as a technique to optimize model performance, efficiency, effectiveness, and uptake. As a result, when done well, participatory AI methods can help to empower historically marginalized or disenfranchised communities and build more trustworthy and accountable AI models.

Academics and experts have proposed a range of methods to advance meaningful public engagement in design and deployment of AI. This includes focus group discussions and structured conversations, surveys, interviews, and citizens' assemblies along with varying approaches to developing AI, such as participatory co-design methods. Despite this plethora of approaches, surveys are increasingly seen as a go-to tool and have been used to both inform and shape the development and deployment of AI, internal policies at AI companies, and national and international policies related to AI.

The increased use of—and perhaps overreliance on—surveys as the dominant approach to participatory AI might be understood and explained by the AI sector's philosophical leanings towards utilitarian ethics and appreciation of quantitative metrics to measure impact. However, critics of surveys note that their use can increase the risks of so-called "participation washing", giving the illusion of participation rather than assigning participants meaningful agency, voice, or power in the decisions related to AI. These risks can be particularly acute for marginalized and disenfranchised groups. Some research suggests that accountability mechanisms can help mitigate the risks of tokenistic participation, particularly where these mechanisms are developed together with communities directly impacted by AI, actively consider the asymmetries of power between communities and those seeking their views, and account for these dynamics in their design.

## 2.3 Humanitarian action and AI

Since the public release of ChatGPT in 2022, humanitarian actors' interest in and engagement with AI tools has increased, in the hopes that these tools will improve the efficiency and impact of their operations, doing more with less against a backdrop of increasingly stretched resources and constrained donor funding. Some research suggests that integrating AI into international aid and organizational structures could yield important productivity gains, with estimates ranging from 14 percent to more than 50 percent.

Humanitarian actors, meanwhile, are trialing a wide range of AI use cases. Sometimes referred to as humanitarian AI, these use cases include rapidly identifying specific geographic areas where need is greatest; improving the management of supply chains and critical aid infrastructure; identifying mortal remains and tracing missing family members; writing proposals for funding; predicting future movements of displaced populations; and anticipating the impact of natural disasters on affected communities.

However, despite this uptick in use, very little has been published on the extent to which crisis-affected communities have been consulted and involved, if at all, in the design and development of humanitarian AI projects. This flies in the face of humanitarians' long-standing commitments to community participation, improving accountability to crisis-affected communities, and shifting power and agency to those communities.

## 3  METHODS

We drew on three analogue methods for this exercise: a comprehensive review of literature; a paper-based, pilot survey; and focus group discussions (FGDs) with participants. Though we recognized the limitations of these methods, we wanted to explore whether we could mitigate them by decolonizing our tools and approach, grounding them in a critically reflexive stance, and decentering our positionality by mutually committing to share power with those engaging in and supporting this pilot public participation exercise.

### 3.1  Location

This pilot public participation exercise was held in December 2024 in 4 locations across Kakuma Refugee Camp and Kalobeyei Integrated Settlement located in northwest Kenya in Turkana County, in close proximity to the borders of Uganda and South Sudan. Originally established in 1992 to host the so-called "lost boys" of Sudan and subsequent influxes of refugees, Kakuma Camp is one of the world's largest refugee camps. Neighboring Kalobeyei Integrated Settlement was established in 2016 to reduce population congestion in Kakuma Camp.

Collectively, Kakuma Refugee Camp and Kalobeyei Integrated Settlement are home to nearly 290,000 refugees and asylum-seekers, with a largely equal split between male and female residents and over 90% originating from four countries: South Sudan (56.6%), Somalia (18%), DR Congo (8.9%), and Burundi (7.1%). Mostly arid and subject to erratic though at times catastrophic rainfalls, this remote portion of Turkana County covers an area of approximately 25 km squared and has a population density of more than 11,000 people per square km—similar to that of Singapore. As in other humanitarian crises, life in the camp is both prison-like and one of exile. Access to vocational training and income-generating opportunities as well as the quality and availability of services—such as health, education, water and sanitation infrastructure—are poor.

Kakuma Camp is divided into four administrative areas—Kakuma 1, Kakuma 2, Kakuma 3, and Kakuma 4—while Kalobeyei Settlement is divided into three villages—Village 1, Village 2, and Village 3. To maximize geographic coverage and participation, this exercise was held in 4 locations: Kakuma 2, Kakuma 4, Kalobeyei 2, and Kalobeyei 3.

### 3.2  Research tools

We developed and drew on the following tools for this pilot public participation exercise: a participation information sheet (PIS), which provided information on the aims of the exercise, background on the CDAC Network and FilmAid Kenya, and the role of participants; an informed consent form (ICF), which included specific information related to participants' privacy and confidentiality, the storage of the survey results and all notes taken, and instances where mandatory referrals to external service providers would be made; a paper-based, pilot survey which included 20 five-point Likert scale questions, ranging "strongly agree" (5) to "strongly disagree" (1); and a FGD guidance note for facilitators which reiterated key points from the PIS and ICF and included seven optional, heuristic questions to help guide FGDs.

### 3.3 Facilitators and participants

FilmAid Kenya staff recruited and remunerated 12 facilitators, 6 men and 6 women, themselves either refugees or Kenyans living in Kakuma or Kalobeyei. Facilitators participated in a one-day training that covered key concepts related to AI, benefits, and risks linked to the use of AI in humanitarian contexts, and community engagement approaches. Facilitators received 4,000 Kenyan shillings (KES) per day for their work, approximately US $30 and about 24% of the average monthly expenditure for a family of five in Kakuma.

Facilitators collectively mobilized 200 individuals across Kakuma Camp and Kalobeyei settlement to participate in the public participation exercise, with the aim of engaging with 50 participants per day over four days. A total of 193 residents presented to participate. Of these 193, survey responses from 1 participant were excluded in the final analysis as he was aged 16 and below the age of consent (18 years) for this exercise. More than 88% of participants hailed from South Sudan (44.8%), DR Congo (27.1%), Ethiopia (9.4%), and Burundi (7.3%). Of the 192 residents who participated in this exercise, 63 (32.8%) were female and 129 (67.2%) were male.

Participants were offered refreshments during the engagement exercise and transport to and from the predetermined locations where the consultations were conducted. Every participant received a stipend of 1,500 KES, approximately US $11 and about 9% of the average monthly expenditure for a family of five in Kakuma.

### 3.4 Pilot survey and focus group discussions (FGDs)

Approximately fifty people participated in the exercise each day. This included attending a group orientation with a question-and-answer session; completing the paper-based pilot survey; and participating in a focus group discussion led by 2 facilitators with an average of 8 participants per focus group. The first ninety minutes of the session were dedicated exclusively to orientating participants to the exercise. This included a brief overview of AI and how it is being used in humanitarian contexts; an in-depth discussion on participants' entitlements, including their right to withdraw at any time from the exercise and their right to anonymity; and completing the ICF in small groups, supported by a facilitator.

The 20-question survey was distributed and completed in paper format and included questions related to digital and AI familiarity, perceptions of AI, humanitarian use cases, accountability within the humanitarian system as a whole, and AI-related accountability and governance. Facilitators provided detailed explanations and translations of each question. Anonymized responses were coded and entered into a cloud-based database.

After completing the survey, participants engaged in a focus group discussion, with an average of 8 people per group, and led by two facilitators. These FGDs were held in a mixture of languages— including English, French, Juba Arabic, Somali, Swahili—and lasted between 60 and 90 minutes. Facilitators took detailed notes in English. Discussions were audio recorded, where consent for audio recording was received by all participants.

### 3.5 Limitations

As we discuss in more detail below, the use of a paper-based survey alongside other methods, has some notable limitations. Language was, perhaps, one of the most critical. Given resource constraints, we were unable to translate the participation tools into multiple languages beyond English. All facilitators were fluent in English and Swahili and all spoke other languages local to their communities of origin, such as French, Juba Arabic, Somali, and Turkana. The facilitators' understanding of AI, its benefits, and its risks as along with the quality of their translations are likely to have thus shaped participants' own understanding and views. Facilitators' engagement skills would have also impacted participation.

Additionally, completing the surveys and the ICF in small groups might have contributed to "group think", leading some participants to self-censor or conform their views to mirror others in the group, yielding to real or perceived pressure or concerns about lower levels of familiarity with AI and other forms of technology.

Finally, given the differences in distribution across gender and country of origin between participants and the wider population (Appendix A.2), the views surfaced in this pilot exercise should not be considered as statistically representative of views in Kakuma or Kalobeyei, though they do reflect a diversity of views held by residents in those locations. Furthermore, participants weren't explicitly informed that they would receive remuneration before they arrived at the exercise. This may have prevented some participants— particularly women and those facing greater socioeconomic hardship—from volunteering.

## 4 RESULTS AND EMERGENT THEMES

### 4.1 Familiarity with and use of technology

More than 93% of participants reported owning a mobile phone. Amongst female participants, 87% reportedly own a mobile phone while 84% reported owning a smart phone, whereas 96% of male participants reported owning a mobile phone and 87% reported owning a smart phone. A majority of participants (78%) reported having regular and reliable access to the internet, with 82% sending text messages regularly and 28% reporting it difficult to use smart phones. Overall, male participants reported having greater access to the internet than women.

Almost 64% of participants had heard of AI before participating in the exercise and almost as many (61%) reported that they could name a few ways in which AI is being used in the world. Fewer than half (47%) of the participants use AI regularly. There was a marked gender divide in regular AI use between men and women, with 51% of male participants and 37% of female participants reported having used AI before.

### 4.2 Perceptions of AI

A large majority of participants (79%) report feeling generally hopeful and positive about AI, with slightly less optimism from women (76%) than men (81%). A similarly high percentage of participants (76%) believe that AI will generate more benefits than harm. These high levels of optimism appear to hold for both those who reported high levels of familiarity with AI and those who reported little to no familiarity.

However, the same level of optimism did not appear in related questions about participants' personal experiences with AI. Only 52% of participants agreed it had been positive and 30% disagreed. This apparent ambiguity in perceptions of AI might be explained by the limited reliability of participants' self-reported knowledge and understanding of AI, or equally, that ambiguous and somewhat contradictory views of AI are linked to participants' sociotechnical imaginaries. The pervasive myth that AI is impartial and neutral might go some way towards explaining optimistic assessments of AI, particularly for populations living in contexts where chronic corruption, injustice, aid cuts, and limited freedom of movement have a decidedly human face.

### 4.3  Sociotechnical imaginaries

Multiple interesting themes emerged during FGDs which offer some insight into how participants both envision the ways in which AI might shape their communities and their future and, equally, how their cultural values, conceptions of power, social hierarchies, community cohesion, and sociopolitical norms inform these AI sociotechnical imaginaries. Multiple participants raised concerns about AI's potential impact on children, worrying that it might "spoil the minds of our children" or undermine parental authority. Some worried that children might adopt a "mindset of not respecting teachers because they get more information from it [AI]." A number raised concerns that AI would undermine societal support for God, if "people believe in AI instead of depending on God."

The inherent links between cultural values and norms and beliefs of AI's potential harms and opportunities suggested by these comments perhaps underscores the importance of value-driven AI design and co-designing AI solutions with cultural domain experts.

### 4.4  Humanitarian AI use cases and potential impact

Participants broadly agreed (76%) that AI will improve the quality of humanitarian services and support, with 86% women and 73% men agreeing. Equally, 73% of participants agreed that AI will make it easier to find trustworthy and reliable information. Support for the use of AI to determine eligibility for aid and services (69%) was perhaps a bit less pronounced. By contrast, 79% of participants thought AI should be used to identify fraud and abuse perpetrated by aid agencies.

In the FGDs, some participants expressed hope and optimism about the ways in which AI could be used to improve healthcare and opportunities for education. Others were hopeful that AI would "generate malaria vaccines and [a cure for] cancer." These optimistic views were tempered by comments from those who worried that AI would "hinder critical thinking" and lead to a "dull generation." Participants also expressed concern about the impact of AI on individual agency, and labor displacement.

### 4.5  Accountability, AI governance, power, and trust

Multiple views on whom should be held to account, who should govern AI, and who might be trusted to use or govern AI were surfaced through both the survey and the FGDs. Opinions on accountability varied: some suggested tech companies and model developers should bear primary responsibility for AI harms; others placed ultimate responsibility on the Government of Kenya and UNHCR as the primary duty bearers in the camp. Some noted the "need to involve the community leaders" while others suggested community-based committees that include those who have undergone AI training.

Overwhelmingly, participants felt strongly that people like them should influence decisions about how AI could be used in crises, with 83% agreeing. In FGDs, some considered participation in AI deployment an effective way to ensure accountability and mitigating risks. "We need to be involved in decisions," said one participant, "because we refugees are the ones affected." Some suggested their values and cultural expertise should influence the design and use of AI. "AI needs to be installed by some cultural information so that it can follow the guidelines of the community."

Participants' strong views about their role in AI governance sits in sharp contrast to their views about their own individual agency and power to influence decisions about humanitarian aid and related services, where only 63% of participants felt they had the necessary information and resources to do so. Moreover, only 55% of participants agreed that they felt that they could correct errors in aid and service delivery, despite decades of targeted work and commitments to improve community participation and accountability by humanitarian actors.

This apparent tension between voice and agency in AI governance versus humanitarian action might be explained by differences in aspirational rather than realized power. The governance of and accountability linked to AI-related projects necessarily sits within the rigid hierarchies of power in Kakuma and Kalobeyei. The Government of Kenya and UNHCR, as well as other humanitarian actors, wield tremendous power—often serving as the final arbiters in life-altering decisions. Where crisis-affected communities might aspire to participate in AI governance, previous experience perhaps demonstrates that their views, if solicited, can be—and seemingly have been—simply ignored.

# 5 DISCUSSION

## 5.1 The hidden harms of surveys

Through this pilot public participation exercise, we found surveys to be an insufficient tool for meaningful engagement with populations as diverse as those in Kakuma and Kalobeyei. Not only can surveys hide or exclude subtle differences in opinion between different groups as well as individual survey responses but they are limited in their ability to offer reasonable or even potential explanations for these differences. In our pilot exercise, the FGDs proved helpful in surfacing nuances in participant views and helping to explore the answers participants gave, though inevitably, the quality of this exploration was highly dependent on the facilitator's individual skills and abilities. Moreover, as a research tool, surveys risk reducing moral values, cultural norms, and personal views to quantitative, tractable assessments which, in a worst-case scenario, might be engineered to give the illusion of consent and representation or used to manufacture consent entirely.

## 5.2 Participation and perceptions of power

We strongly suspect that the power dynamics and differentials—both between researchers and participants and between individual participants—shaped the outputs of the exercise, despite conscious efforts to ground our research in reflexivity processes, decolonize our research, and decenter our positionality. Both FilmAid Kenya and CDAC command resources and wield power in the camp: FilmAid Kenya as a provider of aid and implementing partner of UN agencies and CDAC as a network of aid agencies that includes some of the world's largest aid actors. We must, therefore, question whether some participants felt compelled—either consciously or unconsciously—to share views which they believed aligned with our personal views or agency-specific positions on AI.

Moreover, it is likely that the power differentials between participants—derived from age, gender, ethnicity, educational attainment, and familiarity with AI or other forms of technology, amongst others— played an important role in shaping the FGDs. It is possible that the views of those participants with lower AI familiarity may have been shaped by others in the group who were perceived as "experts" or that they withheld their views altogether.

## 5.3 Domain expertise

Debates around knowledge construction and domain expertise—whose expertise matters, which domains are more or less important, whether expertise can be extracted and coded—and how they inform the design and development of AI models and use cases are longstanding. Research has shown how the specific lived experiences of end users and the expertise of those working in fields outside of computer science are often undervalued or, in some cases, completely overlooked. As noted above, several participants argued that the expertise and inputs of those with lived experience as refugees should influence the design and use of AI. Taken together with the prevailing power imbalances within the humanitarian sector, actions that favor AI domain expertise over the lived experience of crisis-affected communities could increase the risks of tokenistic participation.

## 5.4 A way forward

Our experience suggests that the scope of the permissions afforded by the participatory AI tools we trialed and the context in which we used them could allow some researchers, in some circumstances, to select data or repurpose conversations—either consciously or unconsciously—and generate outcomes which don't necessarily align with community views and values. So, if these approaches are insufficient and potentially risky, where do we go from here?

Meaningful engagement and participatory AI in humanitarian contexts require interventions which go beyond considerations of AI's potential harms and benefits and ones that employ data justice and feminist AI methods to understand how the power of humanitarian actors and AI developers influences the problems they seek to address and the solutions they offer. Where surveys and one-off discussions with communities fall short of meaningful engagement, long-term and deliberative approaches to participatory AI offer hope.

AI-related citizens' councils or assemblies are springing up across Europe and North America. A similar structure in settings like Kakuma and Kalobeyei, with rotating membership, could support the co-design of AI solutions rather than ex-post facto consultations to seek community appropriation or "manufactured consent". Members could receive regular training and education on AI and work with existing community structures to extend knowledge and awareness on AI beyond the council.

To avoid tokenistic participation, this outreach would need to be backed by appropriate governance arrangements that translate the outputs of consultation processes into deliberate actions and measurable pathways for change, directly impacting how AI models are trialed and used in humanitarian contexts. This could include a process that requires aid agencies and those using AI in humanitarian contexts to publicly report, in the name of transparency and accountability, if and how they have addressed or taken up recommendations made by the citizens' council. Both of these processes acknowledge the need to deliberately shift decision-making power and agency to communities directly impacted by AI and the importance of recognizing cultural values and lived experience as "domain expertise" which is as valuable and legitimate as the expertise offered by AI engineers and computer scientists.

## 6 CONCLUSION

In this paper, we have argued that while calls for participatory AI are increasing across the Global North, there has been a notable absence of efforts to engage crisis-affected communities in the design of AI solutions in humanitarian contexts. Increasing public consultations on AI with these communities is not only ethical and in-line with existing commitments humanitarian actors have made related to community participation, accountability, and shifting power to the Global South, but will contribute to more effective and responsible AI solutions and policy.

Participation is not a panacea for responsible AI in high-risk contexts like humanitarian crises. On the contrary, when used without careful thought and design, some qualitative and quantitative participatory AI methods—including surveys and focus group discussions—might generate results which could be interpreted in multiple ways, thus increasing the risks of tokenistic participation. However, in the absence of meaningful participation, AI solutions deployed in humanitarian contexts risk becoming experimental and extractive, generating harm for those they intend to help. Moreover, the structural and historic imbalance of power between crisis-affected communities and aid providers remains a key obstacle to participatory AI in humanitarian crises.

To ethically and responsibly deploy AI in these contexts, aid actors must challenge these inequities of power and commit to recalibrating the relationship with crisis-affected communities whilst building long-term, deep, and purposeful engagement methods, such as citizens' assemblies, and ongoing education and information sharing on AI that transfer knowledge about AI and agency to crisis-affected communities. These methods should ensure broad representation, facilitate open and informed dialogue, and establish transparent and formal feedback loops between communities and those leading AI projects. Equally, aid actors must publish when and how they engage with communities around the use of AI and report on how they have incorporated their views, both as a means of increasing transparency around the use of AI and to strengthen accountability to the communities they serve. If done well, participatory AI in humanitarian contexts might not only strengthen humanitarian action but also improve global AI practices, reinforcing humanitarian accountability and prioritizing community values, priorities, and agency.

# 7 ETHICAL CONSIDERATIONS

We built into this exercise several measures to uphold and prioritize the privacy, confidentiality, and anonymity of all participants, in line with the ACM Code of Ethics and Professional Conduct, the humanitarian principles, and humanitarian ethics. This included a detailed ICF which, at multiple points throughout the document, underscored the right of all participants to remain anonymous and withdraw from the exercise at any time without incurring any penalty, unfair treatment, or harm. This included the right to withdraw their surveys and any information they shared with facilitators for up to one week after the exercise concluded.

We have committed to sharing and discussing with participants the outcomes of this pilot exercise—including this paper—with all the facilitators and individuals who participated in and supported this work. Not only will this potentially reduce the risks of extractivism that often plague surveys and consultations with crisis-affected communities but contributes to the continuous dialogue and feedback loops that are central to FilmAid Kenya's work and values, as well as those of the authors. Feedback on this project is expected to take place in March or April 2025.

As noted above, we provided remuneration for facilitators and participants. Not only was remuneration intended to reimburse costs associated with participation in the exercise—such as time away from work, childcare, or additional transport—but it is central to our ethical principles and commitments to fairness, social beneficence, and respect for agency and autonomy. Recent research suggests that remunerating participants not only supports the inclusion of socioeconomically marginalized individuals but is a moral necessity and central to the ethics of care and social justice.

# 8 AUTHOR POSITIONALITY STATEMENT

The authors of this paper collectively acknowledge our work, our methods, and our analytical approaches are influenced by our backgrounds, including our lived and professional experiences. We come from various countries and bring a variety of professional and lived experience to this work. Three of the authors live in the Global North (the UK and Ireland) and three authors live in sub-Saharan Africa (Lesotho and Kenya); four are of White European descent and two are of African descent. Two of the authors are computer scientists with deep subject-matter expertise in inclusion, fairness, bias, and underrepresentation in AI and machine learning. Four of the authors have each spent more than twenty years working on issues related to humanitarian action and in crises and contexts of forced displacement in, collectively, over 20 countries across the globe.

We acknowledge that our experience cannot be taken as analogous to nor should it supplant the lived experiences of displaced and crisis-affected communities themselves. We are all committed to democratizing AI and amplifying the agency, voice, and views of communities in the Global South in decisions related to AI development and deployment, in line with feminist AI and decolonizing approaches to AI.

# 9 ACKNOWLEDGMENTS

The authors are grateful for the support of all the FilmAid Kenya staff who contributed their time and expertise to this exercise, without which this work would not have been possible. This includes (listed alphabetically by first name): Augustine Polito, Boston Barongo, Charles Amuton, Edwin Githu, Emmanuel Albert, Gabriella Kennedy, Lauren Edukon, Liwuchoo Ukoth Albino, Lochio Willis, Kuchma Ramadan, Sarah Mutheu Nzioka, Steve Kiza, and Susan Apoo. We are also grateful for the support of all the CDAC Network members, particularly the Global Community of Practice on AI and Community

Engagement experts. Above all, we remain indebted to the 193 women and men who participated in this exercise and our 12 facilitators whose experience, energy, and commitment to equitable and accountable technologies were indispensable. This includes (listed alphabetically by first name): Achola Christine Dominic, Ali Osman Abdulle, Aneta Idiongo Loduho, Chizungu Muchumbi Jereme, Dominic Pasquina Abalo, Lochio Willis Paul, Loure Alele Samuel, Lucien Mulenga, Mzungu Rutunganya Marcel, Saido Omar Noor, Susan Apoo Lochio, and one other (JK) from whom we did not receive permission to print their name.